\documentclass[letter]{aa}  
\usepackage{natbib}
\bibpunct{(}{)}{;}{a}{}{,} 
\usepackage{graphicx}
\usepackage{academicons}
\usepackage{float}
\usepackage{flushend}
\usepackage{txfonts}
\usepackage{hyperref}
\usepackage{tablefootnote}
\newcommand{\gcc}{g\,cm$^{-3}$}

\begin{document} 
\authorrunning{Cs. Kiss et al.} 

   \title{Three fast-rotating Jovian Trojans identified by TESS set new population density limits}


   \author{Cs. Kiss\inst{1,2,3}
          \and
          N.~Tak\'acs\inst{1,2,3}
          \and
          Cs.~E.~Kalup\inst{1,2,3}
          \and
          R. Szakáts\inst{1,2}
          \and
          L.~Molnár\inst{1,2,3}
          \and
          E.~Plachy\inst{1,2,3}
          \and
          K.~Sárneczky\inst{1,2}
          \and
          R. Szab\'o\inst{1,2,3}
          \and
          Gy.~M.~Szab\'o\inst{4,5,1}
          \and
          A.~Bódi\inst{1,2,6}
          \and
          A. P\'al\inst{1,2,3}
    }
\institute{Konkoly Observatory, HUN-REN Research Centre for Astronomy and Earth Sciences, Konkoly Thege Mikl\'os \'ut 15-17, H-1121 Budapest, Hungary
           \email{pkisscs@konkoly.hu}
         \and
           CSFK, MTA Centre of Excellence, Budapest, Konkoly Thege Miklós út 15-17, H-1121, Hungary
         \and
             ELTE Eötvös Loránd University, Institute of Physics and Astronomy, P\'azm\'any P\'eter s\'etany 1/A, Budapest, Hungary
         \and
             ELTE Eötvös Loránd University, Gothard Astrophysical Observatory, Szent Imre h. u. 112, 9700, Szombathely, Hungary
         \and
             MTA-ELTE Exoplanet Research Group, Szent Imre h. u. 112, 9700, Szombathely, Hungary
        \and
        Department of Astrophysical Sciences, Princeton University, 4 Ivy Lane, Princeton, NJ 08544, USA
         }

   \date{Received ..., 2024; accepted ..., 2025}

 
  \abstract
   {Here we report on the identification of the three fastest rotating Jovian Trojans with reliable population assignment, using light curve data from the Transiting Exoplanet Satellite Survey mission, also confirmed by Zwicky Transient Facility data. For two of our targets the rotation periods are moderately below the previously accepted $\sim$5\,h Jovian Trojan breakup limit (4.26 and 4.75\,h), however, the rotation period of (13383) was found to be P\,=\,2.926\,h, leading to a density estimate of $\rho$\,$\approx$1.6\,\gcc, higher than the generally accepted $\lesssim$1\,\gcc\, density limit of Jovian Trojans. If associated with lower densities, this rotation rate requires considerable cohesion in the order of a few kPa. The relatively high albedo (p$_V$\,$\approx$\,0.11) and fast rotation suggest that (13383) may have undergone an energetic collision that spun up the body and exposed bright material to the surface.  }

   \keywords{Jovian Trojans --
                TESS --
                Solar System --
                asteroids
               }

   \maketitle
%
\section{Introduction}

Jovian Trojans (JTs) -- asteroids which orbit the Sun at 1:1 mean motion resonance with Jupiter -- are the touchstones of solar system formation and evolution theories. They are {generally} believed to be originated from the trans-Neptunian region, and captured in their current orbits during the giant planet migration era \citep[see e.g.][for recent reviews]{Bottke2023,Mottola2024}. 
{However, some alternative scenarios may be needed to explain the difference between the observed colour distribution of Jovian and Neptune Trojans and trans-Neptunian objects \citep{Jewitt2018}, and in-situ formation \citep{Pirani2019mig,Pirani2019incl} or the effect of stellar flybys \citep[e.g.][]{Ida2000,Pfalzner2024} cannot be completely ruled out.}
There are a number of characteristics of JTs, including surface composition and geology, photometric properties, shape, the existence and properties of satellites and rings, which can provide information on the formation and evolution of these bodies \citep{Marchi2023,Noll2023}. In addition, rotational properties (period and spin axis orientation) provide important constraints on their formation and collisional evolution \citep{Hanus2023,Mottola2024}. 
Assuming a predominantly rubble-pile structure, the rotation period can constrain the bulk density and cohesion necessary to keep the rotating body together by its gravity. In the main belt, this sets the rotational breakup limit of P\,$\approx$\,2.2\,h, which corresponds to a {critical} density of $\rho_c$\,$\approx$\,2.2\,\gcc\, for asteroids larger than a few km in size \citep{Pravec2000}, {Asteroids of this density would disintegrate if they rotated faster. As faster rotating asteroids would require higher densities to remain intact, the breakup limit (spin rate of the fastest rotating asteroids) defines the highest density observed in a specific population}. 
The presently known fastest-spinning Jovian Trojan with a reliable orbit determination is asteroid (187463) 2005~XX$_{106}$, with a rotation period of P\,=\,4.84\,h, identified by \citet{French2015}. 
This rotation period, and those of other JTs with similar spin rates, set a {population} density limit of $\sim$0.9\,\gcc\, assuming ruble-pile structure \citep[see][for a summary]{Mottola2024}.  
This low density of JTs also indicates icy compositions and a considerable level of porosity. 

Here we report on identifying the fast (P\,$\leq$4.8\,h) rotation of three Jovian Trojans using data from the measurements of the Transiting Exoplanet Survey Satellite \citep[TESS,][]{Ricker2015}, the fastest rotators detected in this dynamical population. We present our observations and data reduction in Sect.~\ref{sect:obs}, our results in Sect.~\ref{sect:results}
and discuss the implications of the these newly identified fast rotation rates in Sect.~\ref{sect:discussion}. 


   \begin{figure*}
   \centering
    \includegraphics[width=\textwidth]{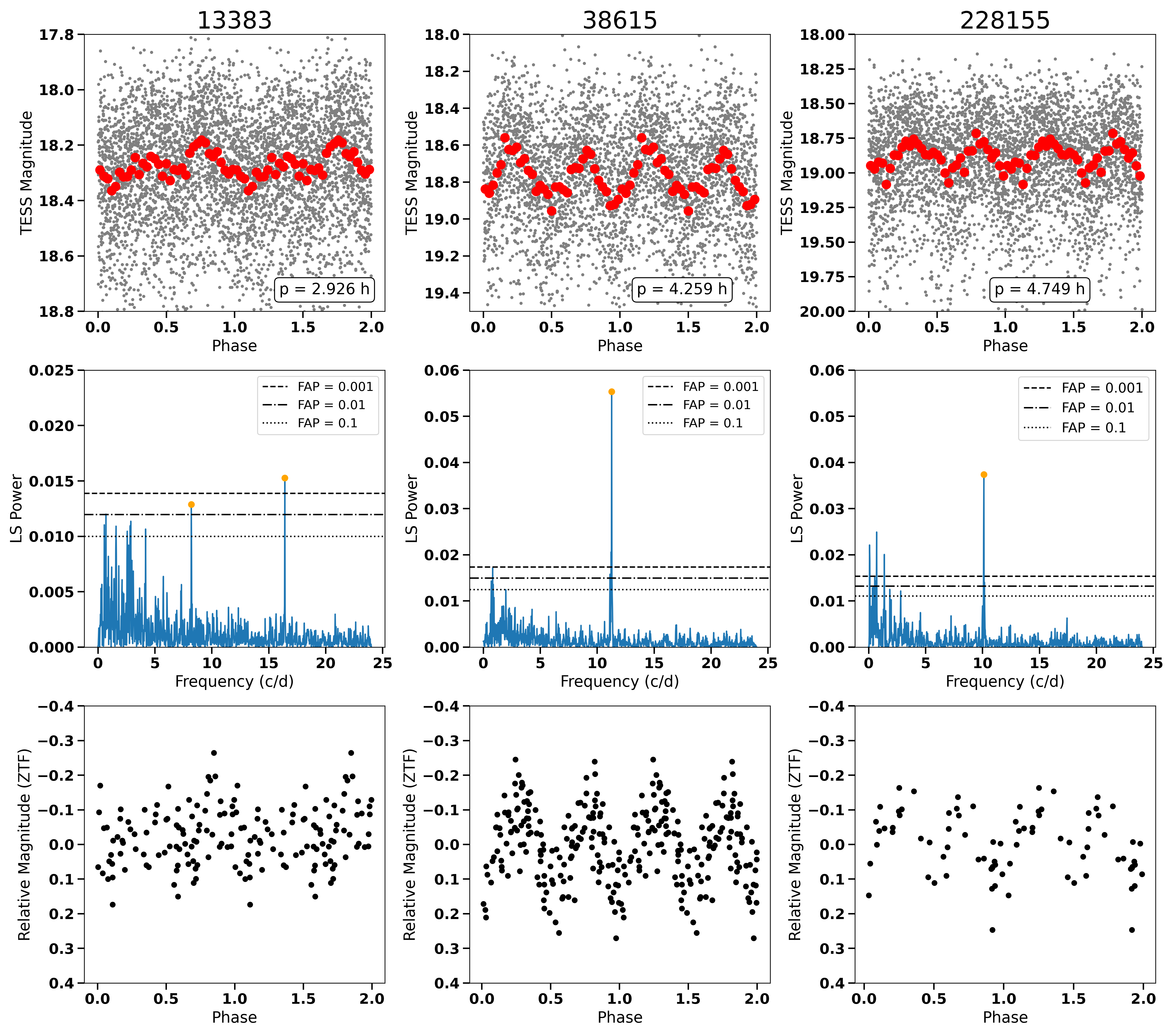}
   \caption{Top: TESS light curves folded with the rotation periods. Grey points are individual observations. We combined the data into 36 phase bins and show the binned rotation curve with large red points. Middle: {power} spectra. For (13383) both the rotation frequency and its double are present. For (38615) and (228155), only the double-frequency (half-period) signal is significant, due to their symmetric rotation curves. Bottom: light curves from the ZTF survey, folded with the same periods. \label{fig:lcs}}  
    \end{figure*}

\section{Observations and data reduction}
\label{sect:obs}

In this paper, we present light curves and photometric properties of three fast-rotating Jovian Trojan asteroids: (13383), (38615) and (288155), observed by TESS during Sectors 29, 42, 43 and 44 (see Table \ref{table:targets}). Data reduction steps are analogous to the first data release of photometric analysis provided by the TESS observations for small solar system bodies (TSSYS-DR1), as discussed in \cite{Pal2020}. While TSSYS-DR1 contains nearly ten thousand light curves and the derived rotation characteristics (such as accurate periods and amplitudes) from Year 1 of the TESS mission, the second data release is processed with higher accuracy and precision in terms of photometric analysis up to the end of Year 4. Such enhancements include the application of a series of elongated apertures -- providing optimal overall signal-to-noise ratios for both fainter and brighter asteroids --, accurate propagation of noise characteristics from background contamination caused by variable stars, and automatic masking of data points for mutual asteroid crossings. Imaging data are processed exclusively by the tasks of the FITSH package \citep{Pal2012}.

During the {post-processing} of the derived light curves, first we eliminated the remaining obvious outliers using iterative sigma clipping with the LOWESS algorithm \citep{Cleveland}. Then, a low-order least-squares polynomial fitting was applied to remove any remaining instrumental trends and phase-angle variations {\citep{Pal2020,Vavilov2025}}. We obtained rotation rates using Lomb-Scargle periodograms, which were validated by visual inspection. We note that for asteroid (228155) we were able to clearly detect the rotational signal not only in Sector 29, but also in Sectors 42 and 43. For asteroid (38615), data from Sector 43 Camera 2 CCD 2 and CCD 3 were combined to derive the rotational period.

We also used sporadic photometry from the Zwicky Transient Facility \citep[ZTF,][]{ZTF}. Asteroid detections were identified, and ZTF data were downloaded using the FINK portal\footnote{https://fink-portal.org/}. The ZTF services offer corrections for phase angle and geometrical effects. We utilized the $HG_1G_2$ models, which include geocentric and heliocentric photometric corrections and describe the phase function with two parameters. This is the default phase correction model provided by the FINK portal.
A total of 90, 155, and 127 data points were downloaded for asteroids (13383), (38615), and (228155), respectively. The observations covered the years 2021–2024, while for asteroid 13383, 9 data points from 2020 were also included. Based on these data and the TESS periods, we plotted the phase diagrams shown in the lowest panels of Fig.~\ref{fig:lcs}.  

\section{Results \label{sect:results}}

The power spectra of the asteroids (38615) and (288155) show a single prominent peak, at f\,=\,11.270 and 10.107\,cycle/day (c/d) frequencies, respectively (see Fig.~\ref{fig:lcs}). When folded with the half frequency/double period the light curves show slight asymmetries in the first and second half periods, therefore we accepted these double-peaked period of P\,=\,4.259$\pm$0.002 for (38615) and P\,=\,4.749$\pm$0.001 for (288155). 


\renewcommand{\arraystretch}{1.3}
\begin{table}[!ht]
\begin{center}
\caption{Main results of the asteroid light curve period determination}
\begin{tabular}[t]{c|ccc}
\hline
\hline
\textbf{ID}          &   \textbf{(13383)}   &    \textbf{(38615)}   &    \textbf{(228155)}   \\ 
                     &       1998 XS3     &      2000 AV121     &       2009 SF61      \\ \hline
\textbf{SCC}         &        S44C1C2     &       S43C2C2/3       &      S29C1C4         \\
\textbf{P (h)}       &  2.926$\pm$0.002  &  4.259$\pm$0.002  &  4.749$\pm$0.001  \\
$\mathbf{\Delta m}$ \textbf{(mag)} &   0.18$\pm$0.03    &    0.35$\pm$0.06    &  0.37$\pm$0.07     \\
$\mathbf{H_V} $ \textbf{(mag)}     &        11.10$\pm$0.15       &         12.45       &        12.65       \\
\textbf{p$_V$ (km)}                    &     0.109$\pm$0.022        &          0.07$^*$        &        0.07$^*$          \\
\textbf{D (km)}                    &       24.27$\pm$0.70         &          16$^*$         &        15$^*$          \\
\hline
\end{tabular}
\end{center}
\tablefoot{The columns of the table are: ID: asteroid number; SCC: TESS Sector, Camera and Chip identifier; P: rotation period; $\delta$m: light curve amplitude; $H_V$: absolute magnitude; p$_V$: geometric albedo; D: estimated effective diameter. For (38615) and (288155) we used the population-average geometric albedo value \citep{Grav2012} to calculate the estimated size (marked by $^*$). }
\label{table:targets}
\end{table}

The light curve of (13383) is sufficiently asymmetric to show both the f\,=\,16.40\,c/d peak, and its half frequency (f\,=\,8.20\,c/d) in the power spectrum, clearly identifying the latter one as the main period. For this target we also checked whether the light curve could be `quadruple-peaked', corresponding to a rotation period of P$_4$\,=\,5.852\,h (the double of P\,=\,2.926\,h, f\,=\,8.202\,c/d) by comparing the binned data of first and second halves of the light curve folded with P$_4$, using two similar statistical methods described in \citet{Pal2016} and \citet{Hromakina2019}. Both methods find that the probability that the two halves of the P$_4$ period light curve are different is p\,$\leq$\,0.25, i.e., we see no indication of a quadruple-peaked light curve within the current uncertainties, and we accept the double-peaked period of P$_2$\,=\,2.926\,h as the rotation period. 
{In Fig.~\ref{fig:lcs}, the significance levels of each power spectrum are defined by false alarm levels corresponding to probabilities of 0.1, 0.01 and 0.001 \citep{Scargle1982}. The false alarm probability (FAP) values in the case of (13383) are 0.00328 for the rotational frequency and 0.00018 for its double frequency.}

We also tested the identified periods using ZTF data. 
The tested periodicity is clearly evident in the cases of (38615) and (228155), confirming the period analysis from the TESS data. Additionally, the overall shape of the light curves is also similar. These consistencies strongly support the results for these two fast-rotating Trojans.  
In the case of asteroid (13383), despite that the visual impression also shows similarities between the TESS and ZTF data, the assumed period could not be unambiguously recovered. 
We note, however, that this ZTF light curve had the lowest 
number of data points and signal-to-noise ratio, and the longest overall time coverage, which probably included very different spin axis aspect angles, decreasing the likelihood of a successful period recovery.

   \begin{figure}
   \centering
    \includegraphics[width=\columnwidth]{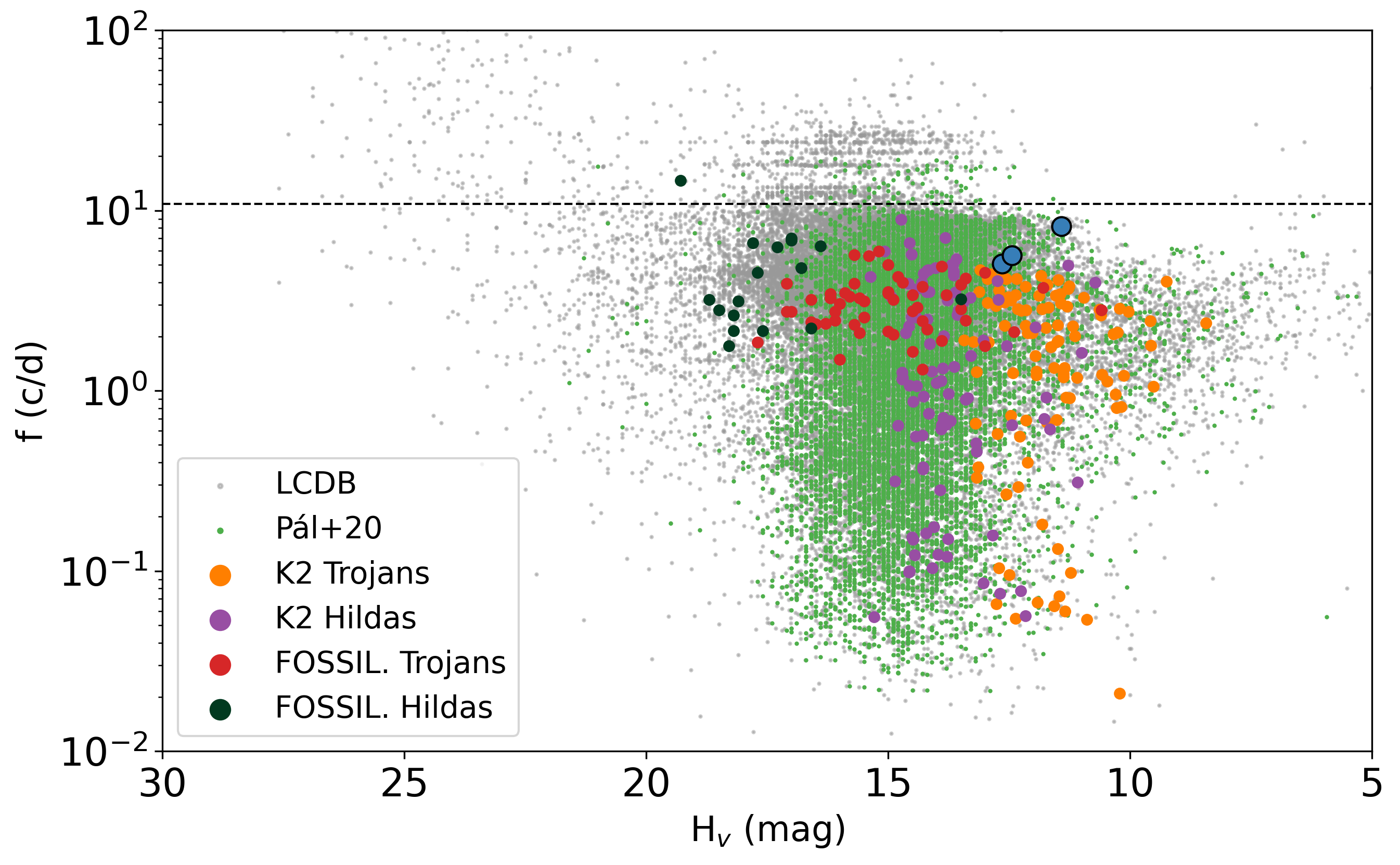}
    \includegraphics[width=\columnwidth]{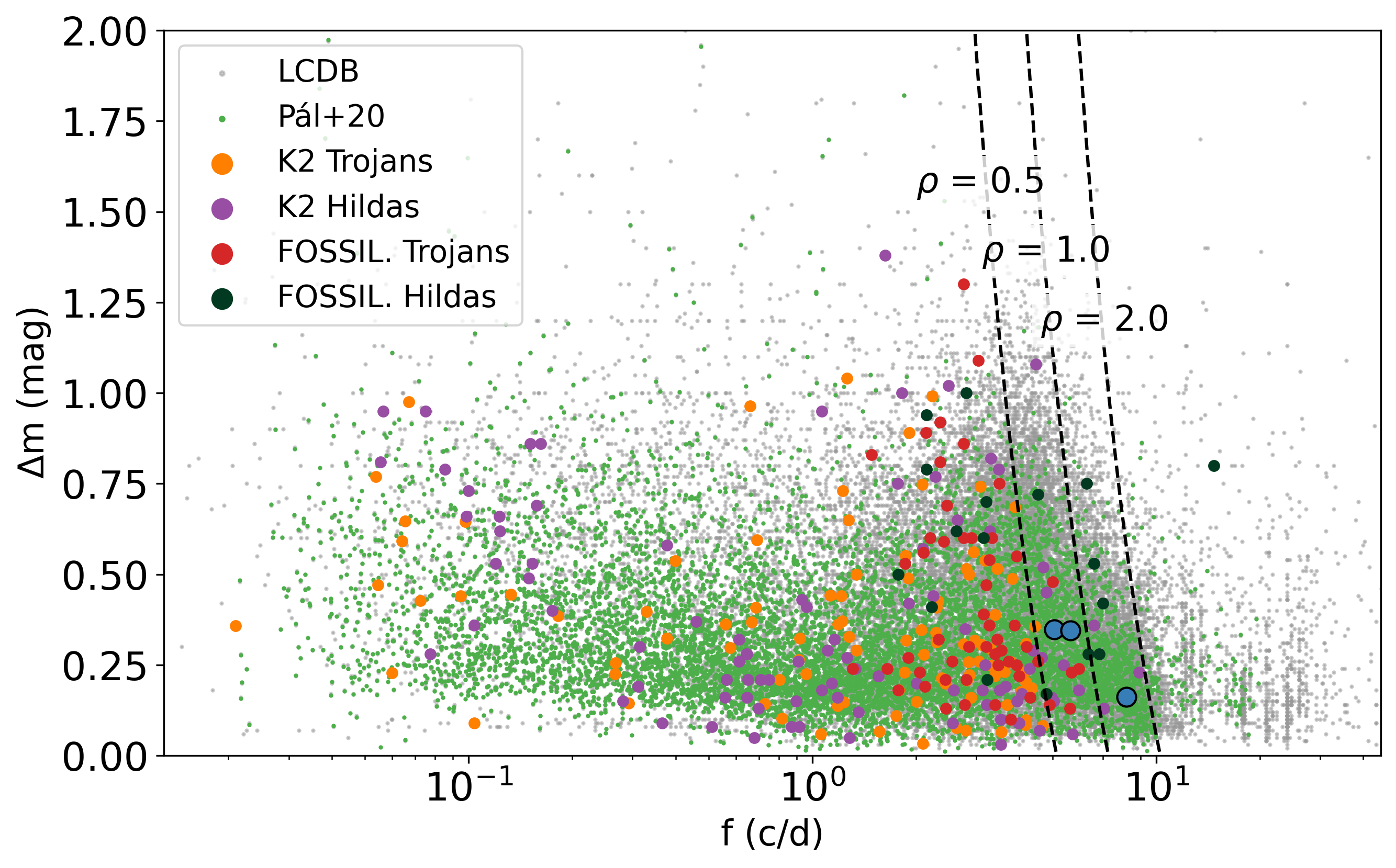}
   \caption{Distribution of basic rotational properties of the inner solar system asteroid populations. Top panel: rotational frequency versus absolute magnitude. The dashed horizontal line marks the breakup limit of main belt asteroids, corresponding to a rotational period of 2.2\,h \citep{Pravec2000}. Bottom panel: light curve amplitude versus rotational frequency. Dashed curves correspond to constant critical densities, as calculated by \citet{Pravec2000}. 
   On both panels, the symbols with different colors correspond to the data sources as marked in the legend box. The three targets discussed in this paper are marked by large blue symbols.}
              \label{fig:freqdistr}%
    \end{figure}

\section{Discussion and conclusions\label{sect:discussion}}

In the FOSSIL survey sample, \citet{Chang2021} found smaller-diameter Jovian Trojans with rotation periods faster than the previously suggested $\sim$5\,h limit, with three of them having rotation periods of $\sim$4\,h. The diameters of these Jovian Trojans are around 5\,km, therefore our targets are the first fast-rotating Jovian Trojans identified in the $\sim$20\,km size range. We also note that \cite{Mottola2024} highlight the desire to confirm these orbital classifications and rotation periods, as they were previously not classified as Trojans, and their assignment is only based on the short-arc determination by FOSSIL.

Assuming a strengthless rubble-pile structure, a lower limit can be estimated for the bulk density from the rotation period and light curve amplitude, following \citet{Pravec2000}. This model, using the previous $\sim$4.8\,h rotation period, led to a density limit of $\sim$0.9\,\gcc\, for Jovian Trojans  \citep{Ryan2017,Chang2021,Mottola2024}. For our targets, the obtained rotation periods and the derived densities are presented in Fig.~\ref{fig:freqdistr}, along with data from large databases, including Jovian Trojans and Hildas from the K2 mission \citep{Szabo2017,Szabo2020,Kalup2021}, and the FOSSIL survey \citep{Chang2021,Chang2022}. The \citet{Pravec2000} relation provides critical densities of $\sim$0.7\,\gcc\, for (38615),  $\sim$0.8\,\gcc\, and (228155), and $\sim$1.5\,\gcc\, for (13383).   

\begin{figure}[ht!]
   \centering
    \includegraphics[width=\columnwidth]{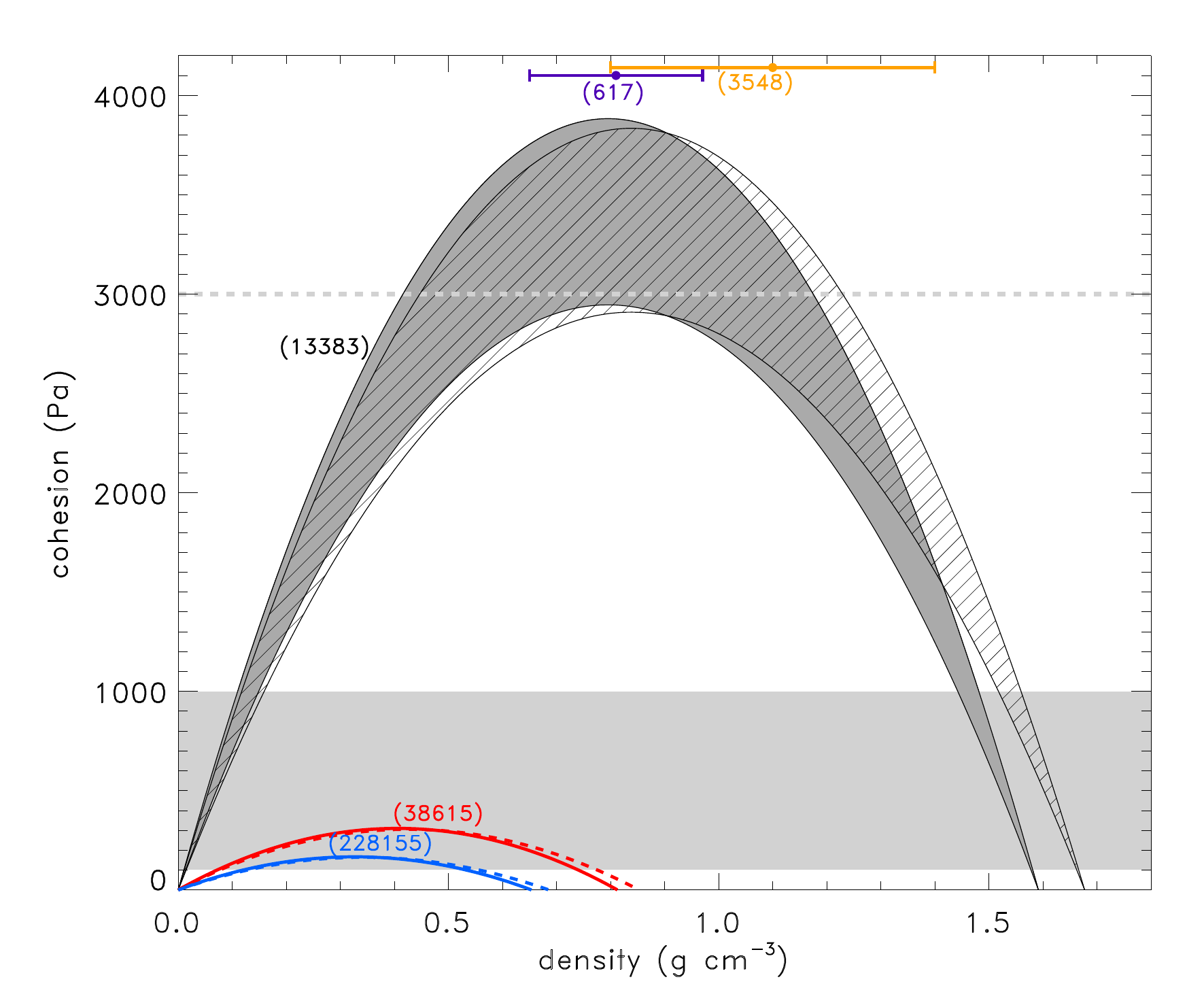}
   \caption{Cohesion versus density curves for our three targets obtained using the Drucker-Prager criterion of failure. Red and blue curves correspond to (38615) and (228155), respectively. {Dashed and solid curves correspond to friction angles of $\phi$\,=\,40 and 45\degr, respectively}.
   The {shaded areas represent} the cohesion curves of (13383) allowed by the uncertainties in absolute magnitude and albedo determination {(striped: $\phi$\,=\,40\degr, solid gray: $\phi$\,=\,45\degr)}. The light gray area represents the typical cohesion range of lunar regolith, while the horizontal dashed line shows the upper limit of the most compacted lunar regolith \citep[$\leq$\,3kPa,][]{Mitchell1974}. Purple and orange symbols represent the densities obtained for the binaries (617) Patroclus-Menoetius and (3548) Eurybates-Queta \citep{Berthier2020,Brown2021}, but without cohesion estimates. }
              \label{fig:stress}%
    \end{figure}

Instead of this strengthless rubble-pile approximation, we may use a simple granular material model. Here we assume that the asteroid has a shape of a triaxial ellipsoid with semi-axes $a\,>\,b\,\geq\,c$, and the rotational breakup limit can be obtained from the Drucker-Prager criterion for failure, following \citet{Holsapple2004,Holsapple2007}. In this model the cohesion necessary to keep the rotating body together is obtained from the criterion $\sqrt{J_2}\,\leq\,k-3sp$, where $J_2$ is the second invariant of deviator stresses, $k$ is the cohesion (shear stress at zero pressure), $s$ is the slope parameter and $p$ is the pressure. An approximate shape of our targets can be obtained assuming that we see the system equator-on, and that the ratio of the $a$ and $b$ semi-axes is simply obtained from observed light curve amplitude {(note that a larger maximum light curve amplitude, or smaller b/a, would lead to a lower cohesion value at a specific density, i.e. this assumption provides an upper limit on the cohesion.)}
Following \cite{Holsapple2007} we also assume that the $b$ and $c$ semi-axes have equal lengths. For (13383) the absolute magnitude, visible range geometric albedo and size are obtained from \citep{Grav2012}. For the other two targets, the effective diameters are obtained from the $H_V$ absolute magnitudes assuming a geometric albedo of $p_v$\,=\,0.07, a typical value among Jovian Trojans \citep{Grav2012} (see also Table~\ref{table:targets}). We assume an angle of friction of $\phi$\,=\,45\degr\, that corresponds to a slope parameter of $s$\,=\,0.356 \citep{Holsapple2007}, {as well $\phi$\,=\,45\degr\ ($s$\,=\,0.315) used in \citet{Polishook2016}, representative of Lunar regolith \citep{Mitchell1974}.}
The results are presented in Fig.~\ref{fig:stress}. 

The densities associated with zero cohesion (k\,=\,0) are similar to those obtained from the strengthless rubble-pile case, {in the case of both friction angles applied} (0.65, 0.8 and 1.6\,\gcc\, for (13383), (38615) and (228155), respectively). 
For (38615) and (228155) the rotation periods confine the possible densities to $\rho$\,$\leq$\,0.8\,\gcc, even in the case of zero cohesion. Lunar regolith has cohesion typically between 
100--1000\,Pa \citep{Mitchell1974}, and main
belt and near-Earth asteroids typically require cohesion in the order of a few hundred Pa \citep[see e.g.][]{Polishook2016}. Assuming these cohesion values would decrease the lower limit on the allowable densities of these two targets. 
For (13383), however, a considerable cohesion is required to allow densities below $\sim$1.5\,\gcc, as presented by the dark gray shaded area in Fig.~\ref{fig:stress}. These cohesion values are notably higher than those of typical lunar cohesion, and may exceed the cohesion of the most compacted lunar regolith \citep[$\lesssim$3\,kPa,][]{Mitchell1974}. 


Reliable density estimates exist for the Jovian Trojan binary systems Eurybates-Queta \citep[$\rho$\,=\,1.1$\pm$0.3\,\gcc,][]{Brown2021} and Patroclus-Menoetius \citep[$\rho$\,=\,0.81$\pm$0.16\,\gcc,][]{Berthier2020}. The likely contact binary (624)~Hektor has ambiguous density values ranging from $\rho$\,=\,1.0$\pm$0.3\,\gcc\, \citep{Marchis2014} to $\sim$2.4\,\gcc\, \citep{Lacerda2007,Decscamps2015}.  
Among the small Kuiper belt objects, (486958) Arrokoth, which has a size of $\sim$18\,km, similar to our targets, has a very low density of 0.16-0.6\,\gcc\, \citep{Keane2022}. However, Arrokoth is a cold classical Kuiper belt object whose formation conditions have likely been different from that of the objects in other, {excited} Kuiper belt dynamical populations, {the likely origin to JTs \citep{Bottke2023}.}
\cite{Grundy2019} finds from the data of binary systems that the densities of D\,$\lesssim$100\,km Kuiper belt objects are $\sim$0.5\,\gcc. From theoretical calculations, \citet{Bierson2019} also estimates densities of 0.5-0.8\,\gcc\, for Kuiper belt objects in this size range, suggesting high porosities. 
While several JTs have densities comparable with those of similar sized Kuiper belt objects, there are some JT systems where the estimated densities definitely exceed these values, including (13383), suggesting that in these cases the link between Kuiper belt objects and JTs may not be straightforward. 

In the discussion above we assumed that our targets are single bodies, however, it is generally accepted that a notable fraction of the JT population can be binaries. Current estimates are mostly based on the fraction of slow rotators \citep{Szabo2017,Ryan2017,Kalup2021} and provide a binary fraction of $\sim$20\%. As it is shown in, e.g., \citep{Lacerda2007} the same system can be modeled assuming a single body (Jacobi ellipsoid) or a Roche binary, and the system, with the actual rotation period, has a significantly smaller rotation parameter $\omega^2/\pi G\rho$\,$\lesssim$\,0.13 than the same system assuming a single body, leading to a higher density in the contact binary case. Similar results have been obtained by \citet{Decscamps2015} assuming dumb-bell-shaped equilibrium figures. In the case of our targets this leads to $\rho$\,$\gtrsim$\,5\,\gcc\ for (38615) and (288155), and an extremely high density of $\rho$\,$\gtrsim$\,13\,\gcc\ for (13383), strongly suggesting that our targets are not contact binaries. 

Jovian Trojans are mostly D-type asteroids, with a smaller number of P, and a small fraction of C-types \citep[see e.g.][]{Barucci2002}. C-type asteroids have typical densities of $\rho$\,$\approx$\,1.5\,\gcc\, in the main belt \citep[see e.g.][]{Vernazza2021}, {and their albedos are typically low, $\overline{p_V}$\,=\,0.066$\pm$0.031 in the D\,$\geq$\,20\,km size range \citet{Usui2013}. The $p_V$\,=\,0.109$\pm$0.022 albedo of (13383) is marginally compatible with the typical albedos of C-type asteroids. }
Dynamically our targets seem to be stable Jovian Trojans, as none of them were identified as `insecure' or `nonresonant' in the recent work by \citep{Greenstreet2024} which aimed to investigate the stability of Jovian Trojans and identify the possible interlopers. 

\citet{Marsett2014} suggested that the high albedo of some Jovian Trojans ({$p_V$\,=\,0.17$\pm$0.07, considering only JTs with reliable albedo estimates in their sample)}
may be due to collision-induced resurfacing, exposing brighter material to the surface. 
{While the albedo a (13383) is {at the lower limit of their high albedo sample}, it is higher than the typical albedo of C-type (as discussed above) or D-type asteroids \citep[$\overline{p_V}$\,=\,0.077$\pm$0.041,][]{Usui2013}. }
As collisions are expected to be the major influence in setting the rotation rates of JTs, rather than the YORP effect at this heliocentric distance \citep{Kalup2021}, {it is still an interesting possibility that} the fast rotation and relatively high albedo of (13383) might be caused by an energetic collision that accelerated the asteroid's rotation to the present spin rate, and notably modified the surface at the same time.

\begin{acknowledgements}
This paper includes data collected by the TESS mission. Funding for the TESS mission is provided by the NASA's Science Mission Directorate. The research leading to these results has received funding from the K-138962, SNN-147362, KKP-137523 and TKP2021-NKTA-64 grants of the National Research, Development and Innovation Office (NKFIH, Hungary). This work made use of Astropy (http://www.astropy.org) a community-developed core Python package and an ecosystem of tools and resources for astronomy \citep{astropy:2013, astropy:2018, astropy:2022}. This work used GNU Parallel \citep{tange2022}. This research made use of NASA’s Astrophysics Data System Bibliographic Services. We thank our reviewer for the useful comments and suggestions. We also thank the hospitality of F\H{o}nix Badacsony where this project was carried out.
\end{acknowledgements}

%
%


\setlength{\bibsep}{1pt}

\bibliography{refs}

\begin{thebibliography}{49}
\expandafter\ifx\csname natexlab\endcsname\relax\def\natexlab#1{#1}\fi

\bibitem[{{Astropy Collaboration} {et~al.}(2022){Astropy Collaboration},
  {Price-Whelan}, {Lim}, {Earl}, {Starkman}, {Bradley}, {Shupe}, {Patil},
  {Corrales}, {Brasseur}, {N{\"o}the}, {Donath}, {Tollerud}, {Morris},
  {Ginsburg}, {Vaher}, {Weaver}, {Tocknell}, {Jamieson}, {van Kerkwijk},
  {Robitaille}, {Merry}, {Bachetti}, {G{\"u}nther}, {Aldcroft},
  {Alvarado-Montes}, {Archibald}, {B{\'o}di}, {Bapat}, {Barentsen},
  {Baz{\'a}n}, {Biswas}, {Boquien}, {Burke}, {Cara}, {Cara}, {Conroy},
  {Conseil}, {Craig}, {Cross}, {Cruz}, {D'Eugenio}, {Dencheva}, {Devillepoix},
  {Dietrich}, {Eigenbrot}, {Erben}, {Ferreira}, {Foreman-Mackey}, {Fox},
  {Freij}, {Garg}, {Geda}, {Glattly}, {Gondhalekar}, {Gordon}, {Grant},
  {Greenfield}, {Groener}, {Guest}, {Gurovich}, {Handberg}, {Hart},
  {Hatfield-Dodds}, {Homeier}, {Hosseinzadeh}, {Jenness}, {Jones}, {Joseph},
  {Kalmbach}, {Karamehmetoglu}, {Ka{\l}uszy{\'n}ski}, {Kelley}, {Kern},
  {Kerzendorf}, {Koch}, {Kulumani}, {Lee}, {Ly}, {Ma}, {MacBride}, {Maljaars},
  {Muna}, {Murphy}, {Norman}, {O'Steen}, {Oman}, {Pacifici}, {Pascual},
  {Pascual-Granado}, {Patil}, {Perren}, {Pickering}, {Rastogi}, {Roulston},
  {Ryan}, {Rykoff}, {Sabater}, {Sakurikar}, {Salgado}, {Sanghi}, {Saunders},
  {Savchenko}, {Schwardt}, {Seifert-Eckert}, {Shih}, {Jain}, {Shukla}, {Sick},
  {Simpson}, {Singanamalla}, {Singer}, {Singhal}, {Sinha}, {Sip{\H{o}}cz},
  {Spitler}, {Stansby}, {Streicher}, {{\v{S}}umak}, {Swinbank}, {Taranu},
  {Tewary}, {Tremblay}, {de Val-Borro}, {Van Kooten}, {Vasovi{\'c}}, {Verma},
  {de Miranda Cardoso}, {Williams}, {Wilson}, {Winkel}, {Wood-Vasey}, {Xue},
  {Yoachim}, {Zhang}, {Zonca}, \& {Astropy Project
  Contributors}}]{astropy:2022}
{Astropy Collaboration}, {Price-Whelan}, A.~M., {Lim}, P.~L., {et~al.} 2022,
  \apj, 935, 167

\bibitem[{{Astropy Collaboration} {et~al.}(2018){Astropy Collaboration},
  {Price-Whelan}, {Sip{\H{o}}cz}, {G{\"u}nther}, {Lim}, {Crawford}, {Conseil},
  {Shupe}, {Craig}, {Dencheva}, {Ginsburg}, {VanderPlas}, {Bradley},
  {P{\'e}rez-Su{\'a}rez}, {de Val-Borro}, {Aldcroft}, {Cruz}, {Robitaille},
  {Tollerud}, {Ardelean}, {Babej}, {Bach}, {Bachetti}, {Bakanov}, {Bamford},
  {Barentsen}, {Barmby}, {Baumbach}, {Berry}, {Biscani}, {Boquien}, {Bostroem},
  {Bouma}, {Brammer}, {Bray}, {Breytenbach}, {Buddelmeijer}, {Burke},
  {Calderone}, {Cano Rodr{\'\i}guez}, {Cara}, {Cardoso}, {Cheedella}, {Copin},
  {Corrales}, {Crichton}, {D'Avella}, {Deil}, {Depagne}, {Dietrich}, {Donath},
  {Droettboom}, {Earl}, {Erben}, {Fabbro}, {Ferreira}, {Finethy}, {Fox},
  {Garrison}, {Gibbons}, {Goldstein}, {Gommers}, {Greco}, {Greenfield},
  {Groener}, {Grollier}, {Hagen}, {Hirst}, {Homeier}, {Horton}, {Hosseinzadeh},
  {Hu}, {Hunkeler}, {Ivezi{\'c}}, {Jain}, {Jenness}, {Kanarek}, {Kendrew},
  {Kern}, {Kerzendorf}, {Khvalko}, {King}, {Kirkby}, {Kulkarni}, {Kumar},
  {Lee}, {Lenz}, {Littlefair}, {Ma}, {Macleod}, {Mastropietro}, {McCully},
  {Montagnac}, {Morris}, {Mueller}, {Mumford}, {Muna}, {Murphy}, {Nelson},
  {Nguyen}, {Ninan}, {N{\"o}the}, {Ogaz}, {Oh}, {Parejko}, {Parley}, {Pascual},
  {Patil}, {Patil}, {Plunkett}, {Prochaska}, {Rastogi}, {Reddy Janga},
  {Sabater}, {Sakurikar}, {Seifert}, {Sherbert}, {Sherwood-Taylor}, {Shih},
  {Sick}, {Silbiger}, {Singanamalla}, {Singer}, {Sladen}, {Sooley},
  {Sornarajah}, {Streicher}, {Teuben}, {Thomas}, {Tremblay}, {Turner},
  {Terr{\'o}n}, {van Kerkwijk}, {de la Vega}, {Watkins}, {Weaver}, {Whitmore},
  {Woillez}, {Zabalza}, \& {Astropy Contributors}}]{astropy:2018}
{Astropy Collaboration}, {Price-Whelan}, A.~M., {Sip{\H{o}}cz}, B.~M., {et~al.}
  2018, \aj, 156, 123

\bibitem[{{Astropy Collaboration} {et~al.}(2013){Astropy Collaboration},
  {Robitaille}, {Tollerud}, {Greenfield}, {Droettboom}, {Bray}, {Aldcroft},
  {Davis}, {Ginsburg}, {Price-Whelan}, {Kerzendorf}, {Conley}, {Crighton},
  {Barbary}, {Muna}, {Ferguson}, {Grollier}, {Parikh}, {Nair}, {Unther},
  {Deil}, {Woillez}, {Conseil}, {Kramer}, {Turner}, {Singer}, {Fox}, {Weaver},
  {Zabalza}, {Edwards}, {Azalee Bostroem}, {Burke}, {Casey}, {Crawford},
  {Dencheva}, {Ely}, {Jenness}, {Labrie}, {Lim}, {Pierfederici}, {Pontzen},
  {Ptak}, {Refsdal}, {Servillat}, \& {Streicher}}]{astropy:2013}
{Astropy Collaboration}, {Robitaille}, T.~P., {Tollerud}, E.~J., {et~al.} 2013,
  \aap, 558, A33

\bibitem[{{Barucci} {et~al.}(2002){Barucci}, {Cruikshank}, {Mottola}, \&
  {Lazzarin}}]{Barucci2002}
{Barucci}, M.~A., {Cruikshank}, D.~P., {Mottola}, S., \& {Lazzarin}, M. 2002,
  in Asteroids III, ed. W.~F. {Bottke}, Jr., A.~{Cellino}, P.~{Paolicchi}, \&
  R.~P. {Binzel} (University of Arizons Press), 273--287

\bibitem[{{Bellm} {et~al.}(2019){Bellm}, {Kulkarni}, {Graham}, {Dekany},
  {Smith}, {Riddle}, {Masci}, {Helou}, {Prince}, {Adams}, {Barbarino},
  {Barlow}, {Bauer}, {Beck}, {Belicki}, {Biswas}, {Blagorodnova}, {Bodewits},
  {Bolin}, {Brinnel}, {Brooke}, {Bue}, {Bulla}, {Burruss}, {Cenko}, {Chang},
  {Connolly}, {Coughlin}, {Cromer}, {Cunningham}, {De}, {Delacroix}, {Desai},
  {Duev}, {Eadie}, {Farnham}, {Feeney}, {Feindt}, {Flynn}, {Franckowiak},
  {Frederick}, {Fremling}, {Gal-Yam}, {Gezari}, {Giomi}, {Goldstein},
  {Golkhou}, {Goobar}, {Groom}, {Hacopians}, {Hale}, {Henning}, {Ho}, {Hover},
  {Howell}, {Hung}, {Huppenkothen}, {Imel}, {Ip}, {Ivezi{\'c}}, {Jackson},
  {Jones}, {Juric}, {Kasliwal}, {Kaspi}, {Kaye}, {Kelley}, {Kowalski},
  {Kramer}, {Kupfer}, {Landry}, {Laher}, {Lee}, {Lin}, {Lin}, {Lunnan},
  {Giomi}, {Mahabal}, {Mao}, {Miller}, {Monkewitz}, {Murphy}, {Ngeow},
  {Nordin}, {Nugent}, {Ofek}, {Patterson}, {Penprase}, {Porter}, {Rauch},
  {Rebbapragada}, {Reiley}, {Rigault}, {Rodriguez}, {van Roestel}, {Rusholme},
  {van Santen}, {Schulze}, {Shupe}, {Singer}, {Soumagnac}, {Stein}, {Surace},
  {Sollerman}, {Szkody}, {Taddia}, {Terek}, {Van Sistine}, {van Velzen},
  {Vestrand}, {Walters}, {Ward}, {Ye}, {Yu}, {Yan}, \& {Zolkower}}]{ZTF}
{Bellm}, E.~C., {Kulkarni}, S.~R., {Graham}, M.~J., {et~al.} 2019, \pasp, 131,
  018002

\bibitem[{{Berthier} {et~al.}(2020){Berthier}, {Descamps}, {Vachier},
  {Normand}, {Maquet}, {Deleflie}, {Colas}, {Klotz}, {Teng-Chuen-Yu}, {Peyrot},
  {Braga-Ribas}, {Marchis}, {Leroy}, {Bouley}, {Dubos}, {Pollock}, {Pauwels},
  {Vingerhoets}, {Farrell}, {Sada}, {Reddy}, {Archer}, \&
  {Hamanowa}}]{Berthier2020}
{Berthier}, J., {Descamps}, P., {Vachier}, F., {et~al.} 2020, \icarus, 352,
  113990

\bibitem[{{Bierson} \& {Nimmo}(2019)}]{Bierson2019}
{Bierson}, C.~J. \& {Nimmo}, F. 2019, \icarus, 326, 10

\bibitem[{{Bottke} {et~al.}(2023){Bottke}, {Marschall}, {Nesvorn{\'y}}, \&
  {Vokrouhlick{\'y}}}]{Bottke2023}
{Bottke}, W.~F., {Marschall}, R., {Nesvorn{\'y}}, D., \& {Vokrouhlick{\'y}}, D.
  2023, \ssr, 219, 83

\bibitem[{{Brown} {et~al.}(2021){Brown}, {Levison}, {Noll}, {Binzel}, {Buie},
  {Grundy}, {Marchi}, {Olkin}, {Spencer}, {Statler}, \& {Weaver}}]{Brown2021}
{Brown}, M.~E., {Levison}, H.~F., {Noll}, K.~S., {et~al.} 2021, \psj, 2, 170

\bibitem[{{Chang} {et~al.}(2022){Chang}, {Chen}, {Fraser}, {Lehner}, {Wang},
  {Alexandersen}, {Choi}, {Granados Contreras}, {Ito}, {Jeongahn}, {Ji},
  {Kavelaars}, {Kim}, {Lawler}, {Li}, {Lin}, {Lykawka}, {Moon}, {More},
  {Mu{\~n}oz-Guti{\'e}rrez}, {Ohtsuki}, {Pike}, {Terai}, {Urakawa}, {Yoshida},
  {Zhang}, {Zhao}, {Zhou}, \& {(The Fossil Collaboration)}}]{Chang2022}
{Chang}, C.-K., {Chen}, Y.-T., {Fraser}, W.~C., {et~al.} 2022, \apjs, 259, 7

\bibitem[{{Chang} {et~al.}(2021){Chang}, {Chen}, {Fraser}, {Yoshida}, {Lehner},
  {Wang}, {Kavelaars}, {Pike}, {Alexandersen}, {Ito}, {Choi}, {Granados
  Contreras}, {Jeongahn}, {Ji}, {Kim}, {Lawler}, {Li}, {Lin}, {Sofia Lykawka},
  {Moon}, {More}, {Mu{\~n}oz-Guti{\'e}rrez}, {Ohtsuki}, {Terai}, {Urakawa},
  {Zhang}, {Zhao}, {Zhou}, \& {Fossil Collaboration}}]{Chang2021}
{Chang}, C.-K., {Chen}, Y.-T., {Fraser}, W.~C., {et~al.} 2021, \psj, 2, 191

\bibitem[{{Cleveland}(1979)}]{Cleveland}
{Cleveland}, W.~S. 1979, Journal of the American Statistical Association,
  74(368), 829

\bibitem[{{Descamps}(2015)}]{Decscamps2015}
{Descamps}, P. 2015, \icarus, 245, 64

\bibitem[{{French} {et~al.}(2015){French}, {Stephens}, {Coley}, {Wasserman}, \&
  {Sieben}}]{French2015}
{French}, L.~M., {Stephens}, R.~D., {Coley}, D., {Wasserman}, L.~H., \&
  {Sieben}, J. 2015, \icarus, 254, 1

\bibitem[{{Grav} {et~al.}(2012){Grav}, {Mainzer}, {Bauer}, {Masiero}, \&
  {Nugent}}]{Grav2012}
{Grav}, T., {Mainzer}, A.~K., {Bauer}, J.~M., {Masiero}, J.~R., \& {Nugent},
  C.~R. 2012, \apj, 759, 49

\bibitem[{{Greenstreet} {et~al.}(2024){Greenstreet}, {Gladman}, \&
  {Juri{\'c}}}]{Greenstreet2024}
{Greenstreet}, S., {Gladman}, B., \& {Juri{\'c}}, M. 2024, \apjl, 963, L40

\bibitem[{{Grundy} {et~al.}(2019){Grundy}, {Noll}, {Buie}, {Benecchi},
  {Ragozzine}, \& {Roe}}]{Grundy2019}
{Grundy}, W.~M., {Noll}, K.~S., {Buie}, M.~W., {et~al.} 2019, \icarus, 334, 30

\bibitem[{{Hanu{\v{s}}} {et~al.}(2023){Hanu{\v{s}}}, {Vokrouhlick{\'y}},
  {Nesvorn{\'y}}, {{\v{D}}urech}, {Stephens}, {Benishek}, {Oey}, \&
  {Pokorn{\'y}}}]{Hanus2023}
{Hanu{\v{s}}}, J., {Vokrouhlick{\'y}}, D., {Nesvorn{\'y}}, D., {et~al.} 2023,
  \aap, 679, A56

\bibitem[{{Holsapple}(2004)}]{Holsapple2004}
{Holsapple}, K.~A. 2004, \icarus, 172, 272

\bibitem[{{Holsapple}(2007)}]{Holsapple2007}
{Holsapple}, K.~A. 2007, \icarus, 187, 500

\bibitem[{{Hromakina} {et~al.}(2019){Hromakina}, {Belskaya}, {Krugly},
  {Shevchenko}, {Ortiz}, {Santos-Sanz}, {Duffard}, {Morales}, {Thirouin},
  {Inasaridze}, {Ayvazian}, {Zhuzhunadze}, {Perna}, {Rumyantsev}, {Reva},
  {Serebryanskiy}, {Sergeyev}, {Molotov}, {Voropaev}, \&
  {Velichko}}]{Hromakina2019}
{Hromakina}, T.~A., {Belskaya}, I.~N., {Krugly}, Y.~N., {et~al.} 2019, \aap,
  625, A46

\bibitem[{{Ida} {et~al.}(2000){Ida}, {Larwood}, \& {Burkert}}]{Ida2000}
{Ida}, S., {Larwood}, J., \& {Burkert}, A. 2000, \apj, 528, 351

\bibitem[{{Jewitt}(2018)}]{Jewitt2018}
{Jewitt}, D. 2018, \aj, 155, 56

\bibitem[{{Kalup} {et~al.}(2021){Kalup}, {Moln{\'a}r}, {Kiss}, {Szab{\'o}},
  {P{\'a}l}, {Szak{\'a}ts}, {S{\'a}rneczky}, {Vink{\'o}}, {Szab{\'o}},
  {Kecskem{\'e}thy}, \& {Kiss}}]{Kalup2021}
{Kalup}, C.~E., {Moln{\'a}r}, L., {Kiss}, C., {et~al.} 2021, \apjs, 254, 7

\bibitem[{{Keane} {et~al.}(2022){Keane}, {Porter}, {Beyer}, {Umurhan},
  {McKinnon}, {Moore}, {Spencer}, {Stern}, {Bierson}, {Binzel}, {Hamilton},
  {Lisse}, {Mao}, {Protopapa}, {Schenk}, {Showalter}, {Stansberry}, {White},
  {Verbiscer}, {Parker}, {Olkin}, {Weaver}, \& {Singer}}]{Keane2022}
{Keane}, J.~T., {Porter}, S.~B., {Beyer}, R.~A., {et~al.} 2022, Journal of
  Geophysical Research (Planets), 127, e07068

\bibitem[{{Lacerda} \& {Jewitt}(2007)}]{Lacerda2007}
{Lacerda}, P. \& {Jewitt}, D.~C. 2007, \aj, 133, 1393

\bibitem[{{Marchi} {et~al.}(2023){Marchi}, {Bell}, {Bierhaus}, \&
  {Spencer}}]{Marchi2023}
{Marchi}, S., {Bell}, J.~F., {Bierhaus}, B., \& {Spencer}, J. 2023, \ssr, 219,
  44

\bibitem[{{Marchis} {et~al.}(2014){Marchis}, {Durech}, {Castillo-Rogez},
  {Vachier}, {Cuk}, {Berthier}, {Wong}, {Kalas}, {Duchene}, {van Dam},
  {Hamanowa}, \& {Viikinkoski}}]{Marchis2014}
{Marchis}, F., {Durech}, J., {Castillo-Rogez}, J., {et~al.} 2014, \apjl, 783,
  L37

\bibitem[{{Marsset} {et~al.}(2014){Marsset}, {Vernazza}, {Gourgeot}, {Dumas},
  {Birlan}, {Lamy}, \& {Binzel}}]{Marsett2014}
{Marsset}, M., {Vernazza}, P., {Gourgeot}, F., {et~al.} 2014, \aap, 568, L7

\bibitem[{{Mitchell} {et~al.}(1974){Mitchell}, {Houston}, {Carrier}, \&
  {Costes}}]{Mitchell1974}
{Mitchell}, J.~K., {Houston}, W.~N., {Carrier}, W.~D., \& {Costes}, N.~C. 1974,
  {Apollo soil mechanics experiment S-200 final report}, {S}pace Sciences
  Laboratory Series 15, Issue 7, Univ. California, Berkeley

\bibitem[{{Mottola} {et~al.}(2024){Mottola}, {Britt}, {Brown}, {Buie}, {Noll},
  \& {P{\"a}tzold}}]{Mottola2024}
{Mottola}, S., {Britt}, D.~T., {Brown}, M.~E., {et~al.} 2024, \ssr, 220, 17

\bibitem[{{Noll} {et~al.}(2023){Noll}, {Brown}, {Buie}, {Grundy}, {Levison},
  {Marchi}, {Olkin}, {Stern}, \& {Weaver}}]{Noll2023}
{Noll}, K.~S., {Brown}, M.~E., {Buie}, M.~W., {et~al.} 2023, \ssr, 219, 59

\bibitem[{{P{\'a}l}(2012)}]{Pal2012}
{P{\'a}l}, A. 2012, \mnras, 421, 1825

\bibitem[{{P{\'a}l} {et~al.}(2016){P{\'a}l}, {Kiss}, {M{\"u}ller},
  {Moln{\'a}r}, {Szab{\'o}}, {Szab{\'o}}, {S{\'a}rneczky}, \& {Kiss}}]{Pal2016}
{P{\'a}l}, A., {Kiss}, C., {M{\"u}ller}, T.~G., {et~al.} 2016, \aj, 151, 117

\bibitem[{{P{\'a}l} {et~al.}(2020){P{\'a}l}, {Szak{\'a}ts}, {Kiss}, {B{\'o}di},
  {Bogn{\'a}r}, {Kalup}, {Kiss}, {Marton}, {Moln{\'a}r}, {Plachy},
  {S{\'a}rneczky}, {Szab{\'o}}, \& {Szab{\'o}}}]{Pal2020}
{P{\'a}l}, A., {Szak{\'a}ts}, R., {Kiss}, C., {et~al.} 2020, \apjs, 247, 26

\bibitem[{{Pfalzner} {et~al.}(2024){Pfalzner}, {Govind}, \& {Portegies
  Zwart}}]{Pfalzner2024}
{Pfalzner}, S., {Govind}, A., \& {Portegies Zwart}, S. 2024, Nature Astronomy,
  8, 1380

\bibitem[{{Pirani} {et~al.}(2019{\natexlab{a}}){Pirani}, {Johansen}, \&
  {Mustill}}]{Pirani2019mig}
{Pirani}, S., {Johansen}, A., \& {Mustill}, A.~J. 2019{\natexlab{a}}, \aap,
  631, A89

\bibitem[{{Pirani} {et~al.}(2019{\natexlab{b}}){Pirani}, {Johansen}, \&
  {Mustill}}]{Pirani2019incl}
{Pirani}, S., {Johansen}, A., \& {Mustill}, A.~J. 2019{\natexlab{b}}, \aap,
  631, A89

\bibitem[{{Polishook} {et~al.}(2016){Polishook}, {Moskovitz}, {Binzel}, {Burt},
  {DeMeo}, {Hinkle}, {Lockhart}, {Mommert}, {Person}, {Thirouin}, {Thomas},
  {Trilling}, {Willman}, \& {Aharonson}}]{Polishook2016}
{Polishook}, D., {Moskovitz}, N., {Binzel}, R.~P., {et~al.} 2016, \icarus, 267,
  243

\bibitem[{{Pravec} \& {Harris}(2000)}]{Pravec2000}
{Pravec}, P. \& {Harris}, A.~W. 2000, \icarus, 148, 12

\bibitem[{{Ricker} {et~al.}(2015){Ricker}, {Winn}, {Vanderspek}, {Latham},
  {Bakos}, {Bean}, {Berta-Thompson}, {Brown}, {Buchhave}, {Butler}, {Butler},
  {Chaplin}, {Charbonneau}, {Christensen-Dalsgaard}, {Clampin}, {Deming},
  {Doty}, {De Lee}, {Dressing}, {Dunham}, {Endl}, {Fressin}, {Ge}, {Henning},
  {Holman}, {Howard}, {Ida}, {Jenkins}, {Jernigan}, {Johnson}, {Kaltenegger},
  {Kawai}, {Kjeldsen}, {Laughlin}, {Levine}, {Lin}, {Lissauer}, {MacQueen},
  {Marcy}, {McCullough}, {Morton}, {Narita}, {Paegert}, {Palle}, {Pepe},
  {Pepper}, {Quirrenbach}, {Rinehart}, {Sasselov}, {Sato}, {Seager},
  {Sozzetti}, {Stassun}, {Sullivan}, {Szentgyorgyi}, {Torres}, {Udry}, \&
  {Villasenor}}]{Ricker2015}
{Ricker}, G.~R., {Winn}, J.~N., {Vanderspek}, R., {et~al.} 2015, Journal of
  Astronomical Telescopes, Instruments, and Systems, 1, 014003

\bibitem[{{Ryan} {et~al.}(2017){Ryan}, {Sharkey}, \& {Woodward}}]{Ryan2017}
{Ryan}, E.~L., {Sharkey}, B. N.~L., \& {Woodward}, C.~E. 2017, \aj, 153, 116

\bibitem[{{Scargle}(1982)}]{Scargle1982}
{Scargle}, J.~D. 1982, \apj, 263, 835

\bibitem[{{Szab{\'o}} {et~al.}(2020){Szab{\'o}}, {Kiss}, {Szak{\'a}ts},
  {P{\'a}l}, {Moln{\'a}r}, {S{\'a}rneczky}, {Vink{\'o}}, {Szab{\'o}}, {Marton},
  \& {Kiss}}]{Szabo2020}
{Szab{\'o}}, G.~M., {Kiss}, C., {Szak{\'a}ts}, R., {et~al.} 2020, \apjs, 247,
  34

\bibitem[{{Szab{\'o}} {et~al.}(2017){Szab{\'o}}, {P{\'a}l}, {Kiss}, {Kiss},
  {Moln{\'a}r}, {Hanyecz}, {Plachy}, {S{\'a}rneczky}, \&
  {Szab{\'o}}}]{Szabo2017}
{Szab{\'o}}, G.~M., {P{\'a}l}, A., {Kiss}, C., {et~al.} 2017, \aap, 599, A44

\bibitem[{Tange(2022)}]{tange2022}
Tange, O. 2022, GNU Parallel 20221122

\bibitem[{{Usui} {et~al.}(2013){Usui}, {Kasuga}, {Hasegawa}, {Ishiguro},
  {Kuroda}, {M{\"u}ller}, {Ootsubo}, \& {Matsuhara}}]{Usui2013}
{Usui}, F., {Kasuga}, T., {Hasegawa}, S., {et~al.} 2013, \apj, 762, 56

\bibitem[{{Vavilov} \& {Carry}(2025)}]{Vavilov2025}
{Vavilov}, D.~E. \& {Carry}, B. 2025, arXiv e-prints, arXiv:2501.07189

\bibitem[{{Vernazza} {et~al.}(2021){Vernazza}, {Ferrais}, {Jorda},
  {Hanu{\v{s}}}, {Carry}, {Marsset}, {Bro{\v{z}}}, {Fetick}, {Viikinkoski},
  {Marchis}, {Vachier}, {Drouard}, {Fusco}, {Birlan}, {Podlewska-Gaca},
  {Rambaux}, {Neveu}, {Bartczak}, {Dudzi{\'n}ski}, {Jehin}, {Beck}, {Berthier},
  {Castillo-Rogez}, {Cipriani}, {Colas}, {Dumas}, {{\v{D}}urech}, {Grice},
  {Kaasalainen}, {Kryszczynska}, {Lamy}, {Le Coroller}, {Marciniak},
  {Michalowski}, {Michel}, {Santana-Ros}, {Tanga}, {Vigan}, {Witasse}, {Yang},
  {Antonini}, {Audejean}, {Aurard}, {Behrend}, {Benkhaldoun}, {Bosch},
  {Chapman}, {Dalmon}, {Fauvaud}, {Hamanowa}, {Hamanowa}, {His}, {Jones},
  {Kim}, {Kim}, {Krajewski}, {Labrevoir}, {Leroy}, {Livet}, {Molina},
  {Montaigut}, {Oey}, {Payre}, {Reddy}, {Sabin}, {Sanchez}, \&
  {Socha}}]{Vernazza2021}
{Vernazza}, P., {Ferrais}, M., {Jorda}, L., {et~al.} 2021, \aap, 654, A56

\end{thebibliography}
\bibliographystyle{aa}


\label{LastPage}
\end{document}